\documentclass{article}

\usepackage{algorithm, algpseudocode}
\usepackage{amsfonts, amsmath, amsopn, amsthm, epsfig}
  \numberwithin{equation}{section}
\usepackage{graphicx, epstopdf}
\usepackage{hyperref}
\usepackage{subfigure}
\usepackage{appendix}
\usepackage{authblk}

\usepackage{calc}
\usepackage[top=2.5cm,bottom=2.3cm,inner=2.3cm,outer=2.3cm,bindingoffset=0.5cm,footskip=0.55cm]{geometry}
\theoremstyle{remark}

\newenvironment{lemma*}[2][Lemma]{\par\bgroup{\bfseries #1\ #2. }\it\ignorespaces}{\egroup}

\title{Toward Speech Separation in The Pre-Cocktail Party Problem with TasTas}

\author[1]{Ziqiang Shi}
\author[2]{Jiqing Han}

\affil[1]{Fujitsu Research and Development Center, Beijing, China}
\affil[2]{Harbin Institute of Technology, Harbin, China}

\newcommand{\BALD}{\begin{aligned}}
\newcommand{\EALD}{\end{aligned}}
\newcommand{\BALDS}{\begin{aligned*}}
\newcommand{\EALDS}{\end{aligned*}}
\newcommand{\BCAS}{\begin{cases}}
\newcommand{\ECAS}{\end{cases}}
\newcommand{\BEAS}{\begin{eqnarray*}}
\newcommand{\EEAS}{\end{eqnarray*}}
\newcommand{\BEQ}{\begin{equation}}
\newcommand{\EEQ}{\end{equation}}
\newcommand{\BIT}{\begin{itemize}}
\newcommand{\EIT}{\end{itemize}}
\newcommand{\BMAT}{\begin{bmatrix}}
\newcommand{\EMAT}{\end{bmatrix}}
\newcommand{\BNUM}{\begin{enumerate}}
\newcommand{\ENUM}{\end{enumerate}}

\newcommand{\BA}{\begin{array}}
\newcommand{\EA}{\end{array}}

\date{}

\begin{document}

\maketitle

\renewcommand{\thefootnote}{\fnsymbol{footnote}}


\begin{abstract}
  In this note,
  we propose to use TasTas~\cite{shi2020speech} for the end-to-end 
  approach to monaural speech separation in the pre-cocktail party problem. 
  Our experiments on the public WSJ0-5mix data corpus results in 10.41dB SDR improvement. 
  If online voice data remixing augmentation~\cite{zeghidour2020wavesplit} is adopted in training, 
  an 11.14dB SDR improvement can be achieved.
  We have open-sourced our re-implementation of the
  DPRNN-TasNet in 
  https://github.com/ShiZiqiang/dual-path-RNNs-DPRNNs-based-speech-separation, 
  and our TasTas
  is realized based on this implementation of DPRNN-TasNet, it is believed that 
  the results in this paper can be reproduced with ease.
\end{abstract}
%
\section{Introduction and Problem Statement}
\label{sec:introduction}

Many methods~\cite{luo2017tasnet,luo2018tasnet,venkataramani2017adaptive,shi2019end, shi2019deep, zhang2020furcanext,luo2019dual,zeghidour2020wavesplit,nachmani2020voice,shi2020speech} 
have shown good performance in two or three speaker separation,
but when we use one of the state-of-the-art methods, TasTas~\cite{shi2020speech}, in the separation of 5 different speakers,
which is a bit like a separation problem in a pre-cocktail party problem, we find 
that there is no naive way to train TasTas successfully at all.
The SI-SDR loss does not decrease at all.  In this note, we try to improve the 
training method of TasTas for 5-speaker separation in a pre-cocktail party problem.
We train each module in TasTas separately. That means we first 
train the ID-Net to extract speaker identity features, then train
the TasNet in the first stage, and finally train the TasNet in the second 
stage to refine the separation results from the first stage.
We will not go into 
the architecture details of 
TasTas~\cite{shi2020speech} again, please refer to the corresponding paper for details.

\section{Speech separation in pre-cocktail party problem with TasTas}
\label{sec:tastasplus}

TasTas does seem to have achieved good performance in the separation of two 
people's voice~\cite{shi2020speech}, but when we use this method for 5 people's voice
separation, the network cannot be trained at all using naive training methods. 
In order to apply TasTas to the speech separation in the pre-cocktail party problem,
we propose a multi-step training method of TasTas. The original 
training algorithm of TasTas is naive, which is direct training.
When TasTas is used for  5 people's voice separation, this training algorithm fails. 
After trial and error, it is found that if we
train different modules of TasTas  separately in steps, it can be trained successfully. The  multi-step training is shown in Algorithm~\ref{alg:TasTasplus},
We first train ID-Net, and then ID-Net will be fixed in the subsequent process. Then we train the first stage of DPRNN-TasNet of TasTas,
and after the training is completed,
this first stage module is also fixed. Finally, we train the second stage refinement 
DPRNN-TasNet of TasTas, until it converges.

\begin{algorithm}[thp]
  \caption{Multi-step training in TasTas}
  \label{alg:TasTasplus}

  1:  Train ID-Net until it converges, and then ID-Net will be fixed in the subsequent process.

  2:  Train the first stage of DPRNN-TasNet of TasTas until it converges, and this DPRNN-TasNet will be fixed in the subsequent process.

  3: Train the second stage refinement DPRNN-TasNet of TasTas until it converges.

\end{algorithm}

For other details of TasTas, e.g. network architecture, training loss, and training methods, please refer
~\cite{shi2020speech} for details.

\section{Experiments}
\label{sec:experiments}

\subsection{Dataset and Neural Network}
\label{ssec:dataset}

We evaluated our system on the 5-speaker speech separation problem using the 
WSJ0-5mix dataset~\cite{nachmani2020voice},
which is a benchmark dataset for 5-speaker mono speech separation in recent years, 
thus most of those methods are compared on this dataset.
WSJ0-5mix contains about 24 hours of training and about 6.3 hours of validation data. 
The mixtures are generated by randomly selecting 49 male and 51 female
speakers and utterances in the Wall Street Journal (WSJ0) training set si\_tr\_s, and 
mixing them at various signal-to-noise ratios (SNR) uniformly
between 0 dB and 5 dB (the SNRs for different pairs of mixed utterances are fixed by 
the scripts provided by~\cite{nachmani2020voice}
for fair comparisons). About 4 hours of evaluation set is generated in the same way, 
using utterances from 16 unseen speakers from si\_dt\_05 and si\_et\_05
in the WSJ0 dataset.

We evaluate the systems with the  SDRi~\cite{fevotte2005bss,vincent2006performance} 
metric used
in~\cite{isik2016single,luo2018speaker, chen2017deep,liu2019divide,wang2019deep}. 
Table~\ref{tab:sdri} lists the
results obtained by TasTas and almost all the results in the past two years, 
where IRM means the ideal ratio mask
\begin{equation}
  M_s=\frac{|X_s(t,f)|}{\sum_{s=1}^{S}|X_s(t,f)|}
  \label{eq:irm}
\end{equation}
applied to the STFT $Y(t,f)$ of $y(t)$ to obtain the separated speech, 
which is evaluated to show the upper bounds of
STFT based methods, where $X_s(t,f)$ is the STFT of $x_s(t)$.

\subsection{Results and Discussions}
\label{ssec:results}

In this experiment,  TasTas is compared with several classical approaches, such as
DPRNN-TasNet~\cite{luo2019dual} and Nachmani's~\cite{nachmani2020voice}. Use 
notation TasTas(I, $x_1$, $x_2$, ... , $x_n$) to
denote our proposed system with  speaker \textbf{I}dentity-aware dual-path  BiLSTM, and $x_1$  
dual-path BiLSTM blocks in the
first stage, $x_2$ blocks in the second stage, etc..
Thus DPRNN-TasNet is just TasTas(6).

Table~\ref{tab:sdri} lists the results obtained by our methods and almost all the results in the past two years,
where IRM means the ideal ratio mask. Compared with these baselines, TasTas obtained an absolute advantage,
once again surpassing the performance of stage-of-the-art. TasTas has achieved the most significant performance improvement
compared with baseline systems, and it breaks through the upper bound of STFT based methods (more than 1.5dB).

For the  \textbf{ablation} study,  Table~\ref{tab:sdri} shows that
 and TasTas(I, 6, 6) is 1.4dB better than TasTas(6) in SDRi. That means the 
iterative multi-phase decontaminated scheme are
effective in boost the performance. As for the ablation study of ID-Net, 
please refer to~\cite{shi2020speech}  respectively.

\begin{table}[th]
  \caption[sdri]{SDRi(dB) in a comparative study of different state-of-the-art
    separation methods on the WSJ0-5mix dataset.}\label{tab:sdri}
  \centering
  \begin{tabular}{c|c}
    \hline
    \hline
    Method                & SDRi  \\
    \hline
    \hline
    IRM                   & 9.6   \\
    DPRNN-TasNet          & 8.4   \\
    Nachmani's            & 10.50  \\
    \hline
    \hline
    TasTas(I, 6, 6)  (ours) & 10.41  \\
    TasTas(I, 8, 9) w/ online remixing~\cite{zeghidour2020wavesplit}  (ours) & 11.14  \\
    \hline
  \end{tabular}
\end{table}

\section{Conclusion}
\label{sec:conclusion}

In this note, we investigated the effectiveness of TasTas~\cite{shi2020speech} 
for 5-talker
monaural speech separation in the pre-cocktail party problem. 
 Benefits from the strength of
end-to-end processing, dual-path BiLSTM, speaker identity consistency loss, 
the multi-stage elaborated iterative scheme, and multi-step training
the best performance of  TasTas achieves 11.14dB SDRi (with 
online voice data remixing augmentation~\cite{zeghidour2020wavesplit}) on the public WSJ0-5mix data corpus.

\section{Acknowledgements}
We would like to thank Dr. Nachmani at Tel-Aviv University \& Facebook AI Research valuable 
discussions on the training of ID-Net.

\bibliographystyle{splncs03}
\bibliography{tastasplus}

\end{document}